\documentclass[aps,pre,final,twocolumn,footinbib,floatfix,twoside]{revtex4}

\usepackage{amsmath,bbm,graphicx,bm}

\newcommand{\deriv}[2]{\frac{\partial #1}{\partial #2}}
\newcommand{\Eq}[1]{Eq.\,\eqref{#1}}
\newcommand{\Eqs}[1]{Eqs.\,\eqref{#1}}

\begin{document}

\title{Event-Driven Dynamics of Rigid Bodies Interacting via
  Discretized Potentials}

\author{Ramses van Zon and Jeremy Schofield} 

\affiliation{Chemical Physics Theory Group, Department of Chemistry,
             University of Toronto, Ontario, Canada M5S 3H6}

\date{\today}

\begin{abstract}%
A framework for performing event\mbox-driven, adaptive time step
simulations of systems of rigid bodies interacting under stepped or
terraced potentials in which the potential energy is only allowed to
have discrete values is outlined. The scheme is based on a
discretization of an underlying continuous potential that effectively
determines the times at which interaction energies change.  As in most
event-driven approaches, the method consists of specifying a means of
computing the free motion, evaluating the times at which interactions
occur, and determining the consequences of interactions on subsequent
motion for the terraced-potential.  The latter two aspects are shown
to be simply expressible in terms of the underlying smooth potential.
Within this context, algorithms for computing the times of interaction
events and carrying out efficient event\mbox-driven simulations are
discussed. The method is illustrated on system composed of rigid rods
in which the constituents interact via a terraced potential that
depends on the relative orientations of the rods.
\end{abstract}

\maketitle

\section{Introduction}

Simulating systems with discontinuous interactions offers a number of
advantages over standard molecular dynamics simulations (SMD) in which
the solution of a system of ordinary differential equations is solved
numerically by iterating a map which approximates the short-time
dynamics\cite{FrenkelSmit}.  The advantage of simulating discontinuous systems using
event-driven algorithms\cite{Rapaport80} (discontinuous molecular dynamics, or DMD)
over SMD is particularly apparent for systems of low density, where
the short-time mapping of the dynamics in SMD is applied to freely
evolve the majority of the system.  In spite of the rarity of
interactions at low density, the inherent time step in SMD
simulations cannot be increased beyond some small threshold without 
loss of stability and accuracy of trajectories, since there is nearly always a small fraction of particles
interacting at any given time.  Several adaptive integration methods
exist, typically based on either separating rapidly and slowly-varying
components of the potential in a multiple time step
approach\cite{mts}, or on time re-parametrization of the Hamiltonian
equations to a new system that is integrated with a fixed step
size\cite{reparam}.  Both methods have inherent drawbacks making them
unsuitable for arbitrary potentials.  On the other hand, in the
event-driven approach where there is no inherent time step, each
non-interacting particle is not propagated forward in time until it
interacts with another particle in the system.

Event driven simulations also have some rather serious drawbacks.  For
example, simulations of flexible molecular systems are plagued with
processing often irrelevant intra-molecular events on very rapid time
scales, wasting a great deal of computation.  When physically
reasonable, much can be gained by treating the molecules as rigid. 
Although the general framework for performing event-driven simulations
of rigid or constrained systems has been recently worked
out\cite{dmdMethod}, numerical methods must be used to find
interaction times, potentially leading to gross inefficiencies.
Another clear drawback that discourages the use of the DMD approach is
that little is known about how to design site-based, stepped
interaction potentials between such rigid bodies, and much work must
be done to tune interaction parameters, such as well-depths and
interaction distances at which discontinuities occur.  The design
of a distance-based, discontinuous potential for long-ranged electrostatic
interactions seems particularly problematic.

In this article, possible solutions for both these problems in DMD
simulations are presented. An algorithm based upon an adaptive grid
search is presented as an alternative to the uniform grid approach of
Ref.~\onlinecite{dmdMethod} to make the search for interaction times
as efficient as possible.   We also demonstrate here how the issue of
designing detailed discontinuous potentials can be side-stepped
altogether by using a mapping of an underlying continuous pair potential onto
discrete potential energy values.  The potential energy then consists
of a set of allowed energy terraces, each mapped onto by many
different positions and orientations of the system.  
The discretization of the potential energy on the level
of the pair potential implies that the evolution consists of free
propagation of the system punctuated by impulses at discrete times
when the underlying continuous interaction potential for a pair of
particles hits a critical value.  Because the scheme uses ordinary
continuous interaction potentials as its basis, no substantial effort
is required to parameterize the Hamiltonian, and the experience of
many years of work in the modeling of systems can be exploited.  The
method is applicable to any type of pair-interaction potential, and
can be used with potentials written for rigid systems that depend on
the center of mass positions as well as the relative orientation of
two interacting bodies\cite{gayBerne,fastWater}.

The paper is organized as follows: Sec.~\ref{System} reviews the
elements of rigid body mechanics required to formulate the method.
In Sec.~\ref{IncludingCollisions}, a general description to
numerically find interaction times is presented. A derivation of the
consequences of the action of the impulsive forces and torques on the
system is given in Sec.~\ref{Rules}.  The scheme is applied to a
simple model system with orientationally-dependent interactions in
Sec.~\ref{modelSystem}. Final comments are given in
Sec.~\ref{Conclusions}.

\section{Rigid Systems Interacting Via Stepped-Potentials}
\label{System}

The systems considered here consist of $N$ rigid bodies, each
of mass $m$ and moment of inertia tensor $\mathsf I_i$.  Associated
with each object $i$ are a center of mass position and velocity
$\mathbf r_i$ and $\mathbf v_i$, an attitude or orientation matrix
$\mathsf A_i$, and an angular velocity vector $\bm\omega_i$.  
The attitude matrix $\mathsf A_i$ transforms a vector $\mathbf a$ with
respect to a fixed inertial lab reference frame to its representation
$\tilde{\mathbf a} = \mathsf A_i \mathbf a$ in the principal-axis
frame of body $i$. $\mathsf A_i$ is in fact defined such that the
moment of inertia tensor is diagonal in the principal axis frame:
$\tilde{\mathsf I}_i = \mathsf A^\dagger_i\mathsf
I_i\mathsf A_i=\mathrm{diag}(I_{i1}, I_{i2}, I_{i3})$, where
$I_{i1}$, $I_{i2}$ and $I_{i3}$ are the (possibly distinct) principal
moments of inertia of body $i$.  Although there are several ways to
parametrize the attitude matrix $\mathsf A$, such as Euler angles,
unit quaternions, or angle-axis\cite{Goldstein} coordinates, three
generalized coordinates are always required to specify the orientation
of each three-dimensional rigid body, denoted here by
$\bm\vartheta_i=(\vartheta_{i1}, \vartheta_{i2}, \vartheta_{i3})$.
The time derivative of $\bm\vartheta_i$ can be related to
$\bm\omega_i$ by noting that $\bm\omega_i$ is related to the time
derivative of the attitude matrix via\cite{VanZonSchofieldCompPhys}
\begin{equation}
  \sum_{a=x,y,z}\varepsilon_{bac}(\bm\omega_i)_a
   = \sum_{a=x,y,z} (\dot{\mathsf A}_i)_{ab} (\mathsf A_i)_{ac},
\label{omegabae}
\end{equation}
where $\varepsilon_{bac}$ is the Levi-Civita symbol.\cite{Goldstein}
From this relation, one can easily derive that $\bm\omega_i=
\mathsf N_i^\dagger\dot{\bm\vartheta}_i$, where $(\mathsf N_i)_{ab} =
\frac12\varepsilon_{bcd}(\mathsf A_i)_{ec}\partial
(\mathsf A_i)_{ed}/\partial(\bm\vartheta_i)_a$.

If the system is governed by a smooth potential $U$, then the
equations of motion imply that
\begin{align}
   \dot{\mathbf p}_i &= - \deriv{U}{\mathbf r_i} = \mathbf F_i;
&
   \dot{\mathbf L}_i &= - \mathsf N^{-1}_i \deriv{U}{\bm\vartheta_i}
                    \equiv \bm\tau_i,
\label{torqueEqn}
\end{align}
where $\mathbf p_i=m\mathbf v_i$ is the (linear) momentum of body $i$,
$\mathbf L_i=\mathsf I_i \bm\omega_i$ is the angular momentum of that
body with respect to its center of mass, $\mathbf F_i$ is the force on
the center of mass of body $i$, and $\bm\tau_i$ is the torque on body
$i$.

In many cases, the formal expression for the torque in \Eq{torqueEqn}
can be written in compact form which does not depend on the choice of
parametrization of the attitude matrix $\mathsf A_i$.  For example, if
the potential can be written in terms of site-site interactions, the
center of mass force $\mathbf F_i = \sum_{\gamma} \mathbf F_{i\gamma}$
is a sum of the forces $\mathbf F_{i\gamma}$ acting on the sites
$\gamma$, and the torque can be written as
\begin{align}
   \bm\tau_i &= \sum_{\gamma} \left( \mathbf r_{i\gamma} - \mathbf r_i \right)
                \times \mathbf F_{i\gamma},
\end{align}
where $\mathbf r_{i\gamma}$ is the position of site $\gamma$ on body $i$.

A second example where one can write simple expressions for $\mathbf
F_i$ and $\bm\tau_i$ is if the potential $U$ is a sum of pair
potentials $U_{ij}$ between bodies $i$ and $j$, each of which is
rotationally invariant but depends on inner products of the relative
position vector $\mathbf r_{ij} = \mathbf r_i - \mathbf r_j$ and a set
of orientationally dependent vectors $\mathbf s_i^\alpha$ and $\mathbf
s_j^\beta$ (where $\alpha$ and $\beta$ are integers indicating
different vectors for body $i$ and $j$, respectively). Then the force
and torque on body $i$ due to interactions with body $j$ through the
interaction potential $U_{ij}(\mathbf r_{ij}, \{\mathbf s_i^\alpha\},
\{\mathbf s_j^\beta\})$ can be written as
\begin{align}
   \mathbf F_{ij}&=-\deriv{U_{ij}}{r_{ij}}\hat{\mathbf r}_{ij} - \sum_{\alpha}
              \deriv{U_{ij}}{(\mathbf r_{ij} \cdot \mathbf s_i^\alpha)}\mathbf s_i^{\alpha}
             - \sum_{\beta}\deriv{U_{ij}}{(\mathbf r_{ij}\cdot\mathbf s_j^\beta)} 
	       \mathbf s_j^{\beta}
\label{Feqn} 
\\
   \bm\tau_{ij} &= - \sum_{\alpha, \beta} \deriv{U_{ij}}{(\mathbf s_i^{\alpha}
       \cdot \mathbf s_j^{\beta})} \, \mathbf s_i^{\alpha} \times \mathbf s_j^{\beta} 
  -\sum_{\alpha} \deriv{U_{ij}}{(\mathbf r_{ij} \cdot \mathbf s_{i}^{\alpha})}
  \, \mathbf s_{i}^{\alpha} \times \mathbf r_{ij}
\label{taui}
\end{align}
where we have used Eq.~({\ref{torqueEqn}) and the fact that $\varepsilon_{bde} (\mathsf N_i)_{ab}
= (\mathsf A_i)_{cd} \, \partial (\mathsf A_i)_{ce} / \partial
(\bm\vartheta_i)_a$.  Using the expressions of the forces in \Eq{Feqn}, one
finds for each interacting pair $\mathbf F_{ij}+\mathbf F_{ji}=0$,
which implies conservation of total linear momentum $\sum_i\mathbf
p_i$.  Furthermore,  using the torques in \Eq{taui}, it is straightforward
to verify that
\begin{align}
   \mathbf r_i \times \mathbf F_{ij} + \mathbf r_j \times
   \mathbf F_{ji} + \bm\tau_{ij} + \bm\tau_{ji} = 0, 
\label{totalTorque}
\end{align}
which, in turn, implies that the total angular momentum $\sum_i (
\mathbf L_i + \mathbf r_i \times \mathbf p_i)$ is conserved by the
dynamics.\cite{fna}

\begin{figure}[t]
\centerline{\includegraphics[scale=0.5]{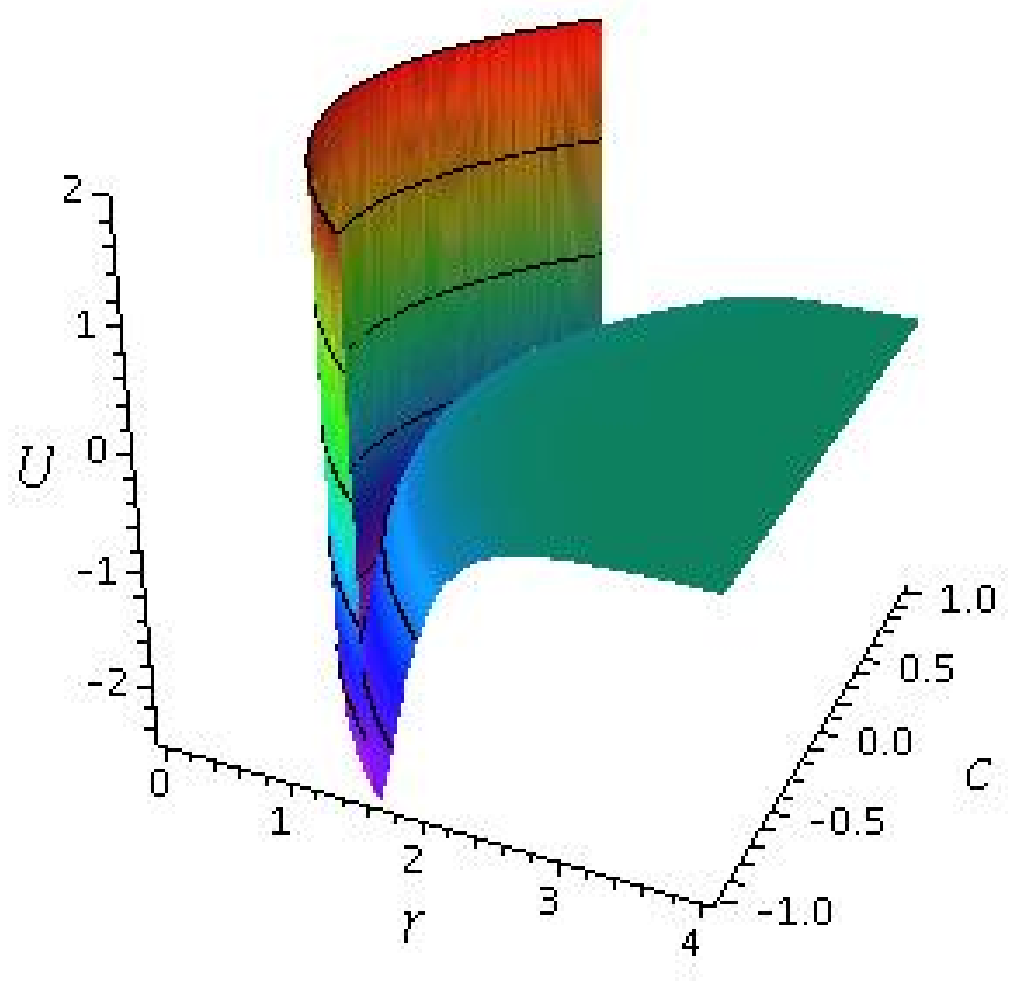}}
\centerline{\includegraphics[scale=0.5]{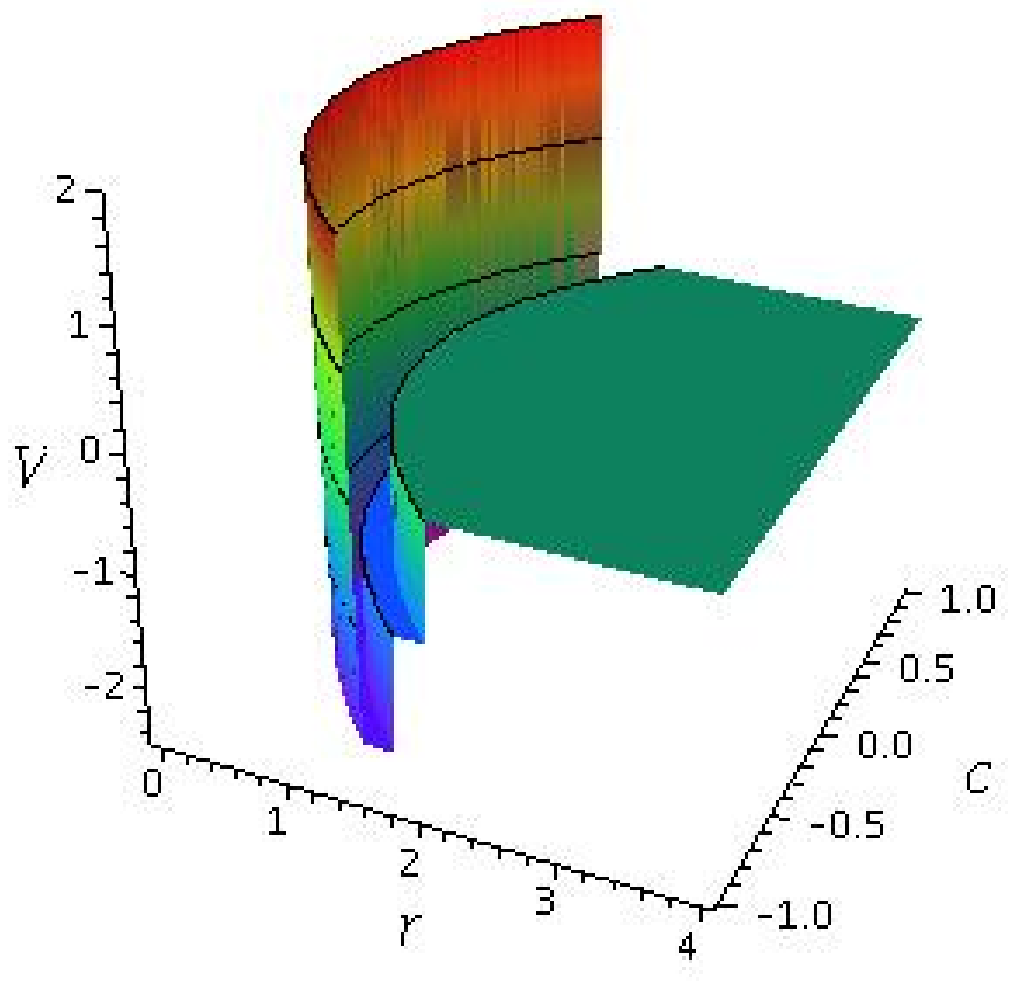}}
 \caption{Illustration of the potential energy discretization
  procedure given in \Eq{steppedPotential} applied to the potential in
  \Eq{RodPotential} of Sec.~\ref{modelSystem}, with the
  $\epsilon=5/2$, $\sigma=1$ and $\Delta=1$. The top panel shows the
  continuous potential $U$ and the bottom panel the terraced variant
  $V$ if $V_{min}=-2$, $\Delta V_k=\Delta V = 1$ and $U_k =
  V_{min}+(k-\frac12)\Delta V$. On the axes, $r=|\mathbf r_i-\mathbf
  r_j|$ and $c=\mathbf s_i\cdot\mathbf s_j$.
  \label{discretize}}
\end{figure}
So far we have considered $U$ to be a continuous potential, which is
required for the derivatives in the above equations to exist.  Now we
consider a stepped interaction potential between a pair $i$, $j$ of
bodies bases on a continuous potential $U_{ij}$, such that the
interaction potential between bodies $i$ and $j$ is of the form
\begin{align}
   V_{ij}&= V_{min} + \sum_{k=1}^K \Theta(U_{ij}-U_k)\Delta V_k,
\label{steppedPotential}
\end{align}
where $\Theta(x)$ is the Heaviside function, and $U_k$ are a discrete
sets of values of the smooth potential at which the system gains or
loses potential energy $\Delta V_k$.  Note that for a system in which
the underlying continuous interaction energy $U_{ij}$ between bodies
$i$ and $j$ lies anywhere in the range $U_k < U_{ij} < U_{k+1}$, the
form of \Eq{steppedPotential} assigns a constant potential
energy value of $V_{min} + \sum_{k' = 1}^{k} \Delta V_{k'}$ to this
interaction.  Figure \ref{discretize} contains an illustration of the
result of this procedure for the potential energy function given in
\Eq{RodPotential}. Because of the shape of the potential energy
landscape, we call a potential of the form \eqref{steppedPotential} a
terraced potential.

For a system in which all interactions are of terraced form, the
procedure to evaluate dynamical properties is the same as that in
simulations of hard sphere systems\cite{dmdMethod}.  While the system
has constant potential energy between interaction events, the motion
of the system is free and the trajectory of each constituent is
independent of all others.  The free propagation of the system between
events determines the evolution of the spatial coordinates of the
molecules in the system, and the interaction times at which the
momenta of the system change discontinuously is determined by
identifying the times at which the underlying continuous pair potential
energy $U_{ij}$ hits an energy terrace $U_k$ where the potential energy
changes discontinuously.  Even though this interaction time is
determined by free motion of the system, in general, it must be found
numerically due to the mathematical complexity of the interaction
condition. The next section addresses this problem. The final
ingredient required to perform event-driven simulations of systems
consists of specifying the interaction rules of how the momenta of the
constituents in the molecular system are altered at the interaction
times and is worked out in Sec.~\ref{Rules}.

\section{Finding Interaction Event Times}
\label{IncludingCollisions}

For systems with terraced potentials as in \Eq{steppedPotential}, the
particles evolve freely until a configuration of the system is reached
where the instantaneous value of the underlying continuous potential
$U_{ij}$ equals $U_k$, for some pair of particles $i$ and $j$.  The
first time at which such an event occurs can be solved by finding the
smallest positive zero (root) of the indicator equation
\begin{align}
  f_{ij}(t) &= U_{ij}\bigl(\mathbf r_{ij}(t), \{ \mathbf s^\alpha_i(t)\},
  \{ \mathbf s^\beta_j(t)\}\bigr) - U_k,
\label{rootEqn}
\end{align}
where the time dependence of the indicator function $f_{ij}$ is
determined by the {\it free} trajectories of bodies $i$ and $j$.  Although
the time dependence of the relative vector $\mathbf r_{ij} (t) =
\mathbf r_{ij}(0) + (\mathbf v_{i} - \mathbf v_{j}) t$ appearing in
the indicator function is simple in the case of free motion, the
time-dependence of the orientational vectors $\mathbf s_i^{\alpha}(t)$
and $\mathbf s_j^{\beta}(t)$ depends on the evolution of the attitude
matrices $\mathsf A_i(t)$ and $\mathsf A_j(t)$, since these vectors
may be expressed as
\begin{align}
  \mathbf s_i^{\alpha}(t) 
  &= \mathsf A^\dagger_i (t)\, \tilde{\mathbf s}_{i}^{\alpha} 
\\
  \mathbf s_j^{\beta}(t) 
  &= \mathsf A^\dagger_j (t)\, \tilde{\mathbf s}_{j}^{\beta},
\end{align}
where $\tilde{\mathbf s}_{i}^{\alpha}$ and $\tilde{\mathbf s}_{j}^{\beta}$ are
time-independent vectors in the body frames of particles $i$ and $j$,
respectively.

The form of the time dependence of the attitude matrices
$\mathsf A_i(t)$ depends on the way in which mass is distributed in
the rigid bodies.  Nonetheless, it is possible to write down
analytical expressions for the time-dependence of $\mathsf A_i$ even
when the mass of a body is not symmetrically
distributed. The general solution of
$\mathsf A_i(t)$ can always be written in the
form\cite{VanZonSchofieldCompPhys}
\begin{align}
   \mathsf A_i(t) &= \mathsf P_i(t) \cdot \mathsf A_i(0),
\end{align}
where the matrix $\mathsf P_i(t)$ propagates the orientation matrix
from an initial time to time $t$. The precise forms of $\mathsf P_i(t)$
for different kinds of rotor can be found in
Ref.~\onlinecite{VanZonSchofieldCompPhys}.

Even with analytical expressions for the time-dependence of the
underlying potential $U_{ij}(t)$, the earliest interaction time must
be found numerically, as was also the case for event-driven dynamics
for rigid systems interacting via site-site
potentials\cite{dmdApplication}.  However, one major benefit of
utilizing an energy terracing approach is that there is only a single
one-dimensional root search to be conducted for each pair of
interacting bodies, in contrast to a site-based energy approach in
which all pairs of sites between pairs of molecules must be examined
for the earliest interaction time.

In Ref.~\onlinecite{dmdMethod}, a simple approach to find the
earliest interaction time was given in which screening methods are
used to identify a minimum and maximum time between which a root of the
indicator function could lie.  This interval is then sub-divided into
equally-sized smaller intervals of fixed size $\Delta t$ and used to
bracket sign changes of the indicator function.  Unfortunately, 
for translating and rotating rigid bodies, the
indicator function is oscillatory, making the detection of so-called
\emph{grazing interactions} troublesome unless a very small grid
interval $\Delta t$ is used.  Since  the
indicator function has a local extremum in a grazing interaction, a good strategy to find this
kind of event is to determine the minimum or maximum of the indicator
function in cases in which the indicator function $f_{ij}(t)$ itself
does not change sign, but its derivative $\dot f_{ij}(t)$
does. However, in the vicinity of a grazing interaction, the values of
the indicator function are typically small, and hence only extrema
where the indicator function at one of the grid points lies below some
threshold value need to be investigated.  Efficient numerical routines
exist to find local extrema of a one-dimensional
function\cite{NumRecipes,Brent}, and thereby detect grazing
interactions. Once an interval with a change in sign of the
indicator function has been identified, standard techniques can be
invoked to find the root\cite{NumRecipes,Brent}.

Although the scheme outlined above is fairly robust and can detect
many millions of events without missing roots, its efficiency is
strongly dependent on the choice of basic time interval $\Delta t$. In
many cases, the size of this interval is determined not by the typical
rate of change of the indicator function but by certain rare scenarios
where rapid changes in the indicator function and its derivative and multiple local extrema occur.  Unnecessarily
small time intervals can be avoided by using the following adaptive
bracketing scheme based on cubic interpolation to estimate the
indicator function and the number of extrema in a given interval.

The basic idea is to use information from previous grid points as much
as possible. Note that when evaluating the indicator function $f$ at a
certain point $t$, it is very easy to also compute its time
derivative.  For a given pair of molecules $i$ and $j$ and assuming a terraced-interaction
potential based on the continuous pair potential $U_{ij}(\mathbf r_{ij}, \left\{
\mathbf s_i \right\}, \left\{ \mathbf s_j \right\})$, the time
derivative of the indicator function $f_{ij}(t)$ is given by
\begin{align}
   \dot{f}_{ij}(t) &= - \mathbf F_{ij} \cdot \mathbf v_{ij} - \bm\tau_{ij}\cdot
   \bm\omega_i - \bm\tau_{ji} \cdot \bm\omega_j,
\label{fdot}
\end{align}
where $\mathbf v_{ij} = \mathbf v_i - \mathbf v_j$ is the relative
velocity vector for the centers of mass, $\mathbf F_{ij}$ is the force
on $i$ due to $j$, and $\bm\tau_{ij}$ and $\bm\tau_{ji}$ are the torques on $i$
and $j$, as given by the smooth interaction in \Eqs{Feqn} and
\eqref{torqueEqn}.

At the start of the search for the first root of the function
$f_{ij}$, its value and derivative are computed. A linear
extrapolation $f_\ell(t) = f_1 + \dot{f}_1 \,t$ can be used to get an
estimate $t'$ for the root of $f_{ij}$ by solving
$f_\ell(t')=0$. The next grid point is then taken to be $t_2=t'+\delta
t$, where a small $\delta t$ is added to enhance the probability of a
sign change of $f_{ij}$ in $[0,t_2]$. At $t=t_2$, the function and its
derivative are evaluated. If there is a sign change in $f_{ij}$, a root has been
bracketed and a numerical root search using standard
techniques is used\cite{NumRecipes,Brent}.
Otherwise, given the value of the indicator function and its derivative at
two times, a unique cubic interpolation of the indicator function can
be constructed.  For example, consider the indicator function $f(t)$
satisfying $f(0) = f_1$, $f(\Delta t) = f_2$ with time derivatives
$\dot{f}(0) = \dot{f}_1$ and $\dot{f}(\Delta t) = \dot{f}_2$.  Using
these values, the cubic approximation $f_{c}(t)$ for the indicator
function over the interval is
\begin{align}
   f_c(t) &= f_1 + \dot{f}_1 \,t + \alpha \, t^2 + \beta \, t^3 ,
\label{cubicInterpolant}
\end{align}
where
\begin{align}
   \alpha &= \frac{3(f_2 - f_1) - (2\dot{f}_1 + \dot{f}_2)\Delta t}{\Delta t^2}
 \\
   \beta &= \frac{2(f_1 - f_2) + (\dot{f}_1 + \dot{f}_2) \Delta t}{\Delta t^3}.
\end{align}
The number of extrema in the interval $[0,\Delta t]$ is then estimated
by finding the number of real roots of the equation $\dot{f}_c (t) =
0$ in the interval.  Furthermore, the values of the extrema are easily
obtained using \eqref{cubicInterpolant}.

\begin{figure}[t]
  \quad\includegraphics[angle=-90,width=0.85\columnwidth]{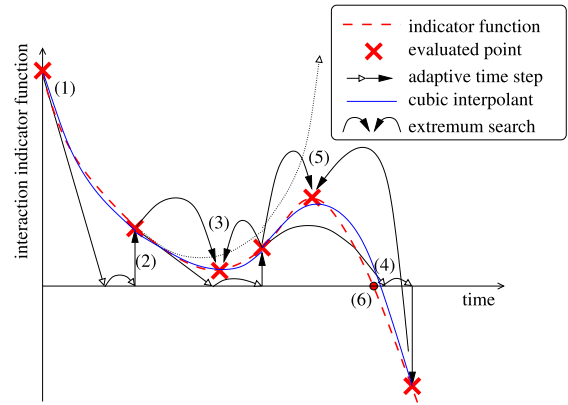}
   \caption{Example of the adaptive root search with cubic
   interpolation.}
\label{rootsearch}
\end{figure}

The adaptive procedure using the cubic approximants is most-easily
explained by considering the example in Fig.~\ref{rootsearch}.
Consider beginning the process of looking for a root after an
interaction event.  In this scenario, at point $(1)$ in the figure,
only the value of the indicator function and its derivative are known
at $t=0$.  The next bracketing point is chosen by solving
$f_\ell(t'_2)=0$.  If the time $t'_2$ is very large or infinite, the
next bracketing time $t_2$ is chosen to be some default (and regular)
value, $\Delta t$.  On the other hand if $t'_2 < \Delta t$, the second
bracketing time $t_2 = t'_2 + \delta t$ is chosen.  At point $(2)$,
the indicator function and its derivative are evaluated and the cubic
and linear approximations to the function evaluated.  The linear and
cubic equations are then examined for roots.  In the case shown in
Fig.~\ref{rootsearch}, the cubic approximation has no roots whereas
the linear interpolant does, at $t'_3$.  Since $t'_3 < t_2 + \Delta
t$, the next bracketing point is selected to be $t_3 = t'_3 + \delta
t$.  But the cubic interpolant has a local minimum in the interval
$[t_2, t_3]$, so a minimum search algorithm is used to find the actual
minimum of $f_{ij}$ at (3).  Since $f_{ij}$ is found to be positive at
(3), a potential grazing interaction has been ruled out. The root of
the cubic $t'_4$ is then used to place the next bracketing point at
$t_4=t_4+\delta t$.  The cubic interpolant using points $t_3$ and
$t_4$ reveals that there is a local maximum in this interval, and that
there is a sign change in the indicator function in the interval since
$f(t_4) < 0$.  The local maximum is investigated and found to be
positive at time $t_5$, so the final bracketing interval where a root
is found is taken to be $[t_5, t_4]$, leading to the root at point
$(6)$.

\section{Interaction rules}
\label{Rules}

If the total interaction potential for the system is a
pairwise-additive combination of terraced two-body interaction
potentials of the form \eqref{steppedPotential}, then forces and
torques only act instantaneously on a pair of molecules $i$ and $j$ at
a time determined by when $U_{ij}$ is equal to one of the reference
values $U_k$.  The forces and torques are therefore impulsive, and
lead to discontinuous changes in the linear and angular momenta of the
pair.  The effect of the impulses must be consistent with conservation
conditions that arise from the symmetries of the overall Hamiltonian,
such as the conservation of energy, and conservation of total linear
and total angular momenta.

Since the momenta and angular momenta change discontinuously and are
not well-defined at an interaction time $t_c$, a more practical
starting point for deriving the interaction rules is to consider the
effect of impulses applied to the overall change in the momentum and
angular momentum of interacting bodies $i$ and $j$ over a small time
interval $[t_c-\delta,t_c+\delta]$:
\begin{align}
   \mathbf p_i' &= \mathbf p_i + \int_{t_c - \delta}^{t_c + \delta} 
                     \mathbf F_i(t) dt 
\label{deltaP} 
\\
   \mathbf L_i' &= \mathbf L_i + \int_{t_c-\delta}^{t_c + \delta} 
                        \bm\tau_i (t) dt,
\label{deltaL}
\end{align} 
with analogous expressions for molecule $j$.  In \Eqs{deltaP}
and \eqref{deltaL}, the primed and unprimed vectors represent post and
pre-interaction values of the respective quantities.  For
small enough $\delta$, the probability of a particle other than $j$
interacting with $i$ becomes zero.  From Hamilton's equations for the
discontinuous system, the impulsive forces and torques are then given by
\begin{align}
   \mathbf F_i (t) &= S^f_i \mathbf F_{ij}(t_c) \delta (t - t_c) 
\\
   \bm\tau_i (t) &= S^l_i \bm\tau_{ij}(t_c) \delta (t-t_c),
\end{align}
where $\mathbf F_{ij}(t_c)$ and $\bm\tau_{ij}(t_c)$ are the forces and torques
on body $i$ at the interaction time $t_c$ for the {\it continuous}
system given in \Eqs{Feqn} and \eqref{taui}, and $S^f_i$ and
$S^l_i$ are unknown scalars.  Note that the directions of the
impulsive forces and torques due to the pair interaction are along the
directions of those quantities for a continuous system interacting by
the same potential at the interaction time.  For the interaction pair
$i$-$j$, conservation of linear momentum immediately implies that
$S^f_j = S^f_i = S^f$. The requirement of conservation
of total angular momentum furthermore gives
\begin{multline}
(S_i^l - S_j^l) \sum_{\alpha, \beta} \deriv{U_{ij}}{(\mathbf s_i^\alpha
 \cdot \mathbf s_j^\beta)} ( \mathbf s_i^\alpha \times \mathbf s_j^\beta )  \\ 
 + (S_i^l - S^f) \sum_{\alpha} \deriv{U_{ij}}{(\mathbf s_i^\alpha \cdot
 \mathbf r_{ij})} ( \mathbf s_i^\alpha \times \bm r_{ij} )  \\ 
 + 
(S_j^l - S^f ) \sum_{\beta} \deriv{U_{ij}}{(\mathbf s_j^\beta \cdot
 \mathbf r_{ij})} ( \mathbf s_j^\beta \times \bm r_{ij} ) = 0.
\end{multline}

Since this condition must be satisfied for arbitrary vectors $\mathbf
r_{ij}$, $\mathbf s_i^\alpha$, $\mathbf s_j^\beta$, we conclude that
$S_i^l = S_j^l = S^f = S$.\cite{fnb} The unknown scalar $S$ is now
found from conservation of energy, by solving the quadratic equation
\begin{multline}
 \biggl( \frac{\mathbf F_{i} \cdot \mathbf F_{i}}{m} + \frac{\bm\tau_i
\cdot \mathsf I_i^{-1} \cdot \bm\tau_i}{2} +  \frac{\bm\tau_j
\cdot \mathsf I_j^{-1} \cdot \bm\tau_j}{2} \biggr)S^2 
 \\
+  ( \mathbf v_{ij}  \cdot \mathbf F_{i} + \bm\tau_i \cdot
\bm\omega_i + \bm\tau_j \cdot \bm\omega_j )S 
+ \Delta V = 0,
\label{quadratic}
\end{multline}
where all quantities are evaluated at interaction time $t_c$ and
$\Delta V$ is the change in potential energy.  Note that the term
proportional to $S$ can also be written as $-\dot{U}_{ij}(t_c)$, as
follows from \Eq{fdot}.  The physical solution corresponds to the
positive (negative) root branch if $\dot{U}_{ij}(t_c) < 0$
($\dot{U}_{ij}(t_c)>0$), provided real roots exist.  If this latter
condition is not met, there is not enough kinetic energy to overcome
the discontinuous barrier, and the system experiences a reflection
with no change in potential energy, i.e., $S$ is the non-zero solution of
\Eq{quadratic} with $\Delta V=0$.

\section{Model System}
\label{modelSystem}

As an illustration of the method, consider a system composed of rods
in which the continuous interaction potential between a given pair
$i$-$j$ is of a modified Lennard-Jones form
\begin{align}
U_{ij}
&= 4 \epsilon
\left[ \left( \frac{\sigma}{r_e} \right)^{12} - \left(
\frac{\sigma}{r_e} \right)^6 \right] 
\label{RodPotential}
\end{align}
with
\begin{align}
r_e &= | \mathbf r_{ij} | + \zeta \left[ \frac{1}{2} - ( \mathbf s_i\cdot
\mathbf s_j )^{2} \right].
\end{align}
Here, $\mathbf s_i$ is a vector pointing along the long axis of the
rod and $r_e$ defines an orientation dependent effective distance of
interaction.  Note that the form of this potential makes it
energetically favorable to align adjacent rods orthogonal to one
another.

By construction, the potential is invariant under
rotations and translations and therefore the dynamics conserves the
total linear and angular momentum in addition to the total energy of
the system.  Using \Eqs{torqueEqn}, \eqref{Feqn} and \eqref{taui}, the force
and torque on body $i$ due to interactions with $j$ is
\begin{align}
   \mathbf F_{ij} &= - \frac{dU_{ij}}{dr_e} \hat{\mathbf r}_{ij} 
\nonumber \\
   \bm\tau_{ij} &= 2\zeta(\mathbf s_i \cdot \mathbf s_j) \frac{dU_{ij}}{dr_e}\,
   \mathbf s_i \times \mathbf s_j 
\nonumber \\
   \frac{dU_{ij}}{dr_e} &= - \frac{24 \epsilon}{r_e} \left[ 2
   \left( \frac{\sigma}{r_e} \right)^{12} - \left(
   \frac{\sigma}{r_e} \right)^6 \right] .
\end{align}
Note that the torque on $i$ is orthogonal to the vector $\mathbf s_i$, so
that the component of the angular momentum along the axis of the
molecule is constant.  Thus, even though the rotational dynamics of
the rigid rods is that of a symmetric top, the angular motion around
the long axis is decoupled from the other directions and can be
neglected. 

From this continuous system one can construct a system with a terraced
potential.  In the current study, the simplest form of the terracing procedure
will be used, where the discontinuities in the potential are
evenly-spaced values of the interaction energy, i.e. in
\Eq{steppedPotential} all $\Delta V_k$ are taken to be equal and the
values of $U_k$ are taken to be $U_k=(k-\frac12)\Delta V$.  In
practice, it is desirable to map a value of $U_{ij} \approx 0$ to
zero, so that pairs that are far apart do not contribute potential
energy to the system.  This is readily accomplished by adjusting the
value of $V_{min}$ in \Eq{steppedPotential}.  It is also important
that the critical values $U_k$ not coincide with minima of the smooth
pair potential, because a terrace of width zero would result, leading to
numerical instabilities.

It is interesting to consider how the graininess of the energy
terraces influences dynamical phenomena, particular in dense systems
where orientational relaxation times become quite long.  To examine
the sensitivity of dynamical correlations to the form of the
stepped-potential, event-driven simulations of a system of $N=512$
rods at a reduced number density $\rho^* = \rho \sigma^3 = 0.5$ and
reduced temperature $T^* = kT/\epsilon = 1.4$ were carried out in a
periodic, cubic box and contrasted with SMD simulations using the
continuous underlying potential, \Eq{RodPotential}.  The parameters
chosen for the system and pair potential were $\epsilon = 3.0$ kJ/mol,
$\sigma = 3.035$ \AA, $\zeta = 1$ \AA, $m = 18$ g/mol, and $I_1 = I_2
= 1.154$ g/mol-\AA${}^2$.

\subsection{Implementation of the event-driven simulation}

As was observed in the case of event-driven simulations based on
site-site potentials\cite{dmdApplication}, the efficiency of the
event-driven approach can be greatly enhanced by implementing several
simple and now fairly standard techniques, such as cell divisions,
tree data structures to manage events, and local molecular
clocks\cite{Rapaport80}. 
In Ref.~\onlinecite{dmdMethod}, two additional
techniques for improving the performance of event-driven simulations
utilizing numerical methods of finding event times were also
presented: the use of screening methods to identify initial and final
bracket times, and the truncation of potentially non-useful event
searches through the scheduling of a virtual interaction.  To identify
which pairs of particles could have interactions under the terracing
potential for a given choice for the set of discontinuities, the
maximum distance $r_m$ at which $U$ is larger than $\Delta V/2$ was
found by solving the equation $U_{ij}=-\Delta V/2$ with
$r_e=r_m-\zeta/2$, giving
\begin{align}
  r_m = \frac12\zeta+
  \frac{\sigma}{\sqrt[6]{\frac12[1-\sqrt{1-\Delta V/(2\epsilon)}]}}.
\end{align} 
The system is then partitioned using this maximum interaction distance
to determine the cell size so that only particles in the same or adjacent
cells interact.  For each pair of possibly interacting particles $i$
and $j$, the times at which the distance between center of masses
reaches $r_m$ is solved exactly using the linear free motion of the pair to
determine minimum and maximum bracketing times.

\begin{figure}[t]
   \centerline{\includegraphics[width=0.81\columnwidth]{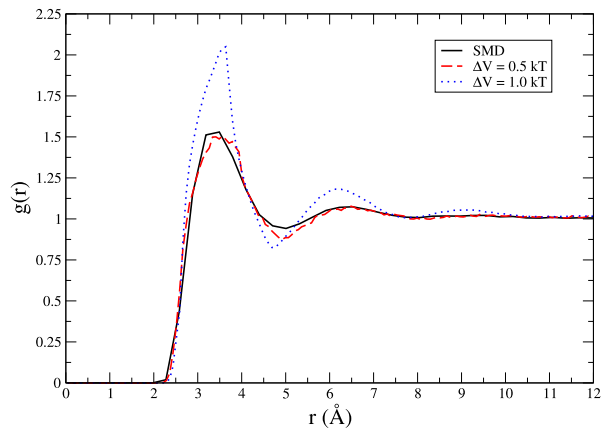}}
   \caption{The radial distribution function for the center of mass of
   the rods.}
\label{rdfFig}
\end{figure}

To implement the virtual interaction event, a root search procedure for
a given pair of rods was only conducted up to a maximum of $4$
femtoseconds before the root search was truncated, and if no root was
found, the value of the indicator function and its derivative at the
end point of the interval were stored in the tree to resume the search
for the pair if neither particle has an event before the point at
which the search was halted.  A maximum step size of $\Delta t = 4$
femtoseconds was also utilized in the adaptive search algorithm
outlined in Sec.~\ref{IncludingCollisions}.

The use of a uniform maximum step size generally leads to sub-optimal
tree structures in the priority queue, where long linear branches
corresponding to scheduled simultaneous virtual interaction events lead
to longer insertion times of new events into the queue.  Two
countermeasures were employed to improve the management of the event
queue:  First, the maximum step size $\Delta t$ was given a small random
adjustment to avoid the coincidence of any two virtual interactions.
The effect of this adjustment is to form more balanced binary trees
free of long linear branches.  Second, a bounded priority
queue\cite{Paul07} was used in which arrays of linear lists of events in a
given time interval are combined with an implicit heap binary tree
that performs a fine sort of the list containing the current time 
interval.  The use of the linear lists enables rapid insertion of
future events into the priority queue, and typically reduced the
overall computational time by around $15\%$ for the system sizes
investigated.  It is expected that the advantage of the bounded
priority queue over standard binary tree data structures will increase
with increasing system size, since operations on the queue take $O(1)$
time per event, as opposed to $O(\log N)$ time.

To keep the tree structure from growing too large in the course of a
simulation, standard scheduling schemes require canceling each event
which becomes invalid due to the occurrence of an earlier interaction
involving one of its participants.  Tracking down these events in the
tree requires a fair amount of bookkeeping\cite{Rapaport80}.  Directly
canceling invalidated events could be done with the bounded priority
queue as well, but there is no real need to do so since the tree
component of the event queue is small already. Not removing invalid
events from the queues was found to lead to another 15\% reduction in
computational time. The problem of growing linear lists was handled by
a (relatively fast and infrequent) cleanup of invalid events from the
lists when computer memory threatens to be depleted.

\subsection{Comparison between standard molecular dynamics and 
event-driven simulations}

To assess the relative merits of the DMD approach as opposed to 
a standard rigid body molecular dynamics approach,
SMD simulation were carried out based on the pair potential in
\Eq{RodPotential}.  The equations of motion were integrated using a
symplectic integration scheme that utilizes the exact free motion for
both rotational and translational degrees of
freedom\cite{integratorPapers}.  The time step used in the SMD
simulations was $1.8$ femtoseconds, resulting in relative fluctuations
of the total energy to the potential energy of about 1\%.  To improve
the efficiency of the SMD simulations,
Verlet lists were used.\cite{FrenkelSmit} In addition, the potential
was smoothly interpolated to zero starting from a cut-off distance of
$r_l = 2.5 \sigma$ and reaching zero at a distance of $r_u = 3\sigma$,
by taking as the interaction potential
\begin{align}
   U_{ij}^{\rm{smooth}} &= g(r_{ij}) U_{ij}(\mathbf r_{ij},
                           \mathbf s_i, \mathbf s_j) 
\label{smoothedPot}
\end{align}
where
\begin{align}
   g(r) &= \left\{
   \begin{array}{ll}
   1 & r < r_l \\
   \frac{(r_u-r)^2(r_u - 3 r_l + 2r)}{(r_u - r_l)^3} & r_l \leq r \leq
   r_u \\
   0 & r>r_u
   \end{array}
\right. .
\label{smoother}
\end{align}

The effect of discretization in the energy-terrace model
on static structural and dynamical correlations was examined by
varying the energy difference
between discontinuities from a fraction of $kT$, the
natural energy unit of the system, to several multiples of $kT$.  Not
surprisingly, the radial distribution function for the center of mass
of the rods, shown in Fig.~\ref{rdfFig}, shows the same qualitative
behavior as that obtained from the continuous potential system, with
clear discretization effects visible when large energy gaps
(i.e. $\Delta V = kT$) are used.  As observed in other systems with
discontinuous potentials\cite{dmdApplication}, the peaks and troughs
observed in the radial distribution function tend to be exaggerated,
but generally in a manner that results in the correct integrated
number of neighbors for each radial shell.  The radial distribution functions
from the SMD and DMD simulations are in excellent agreement when $\Delta V$ is 
$kT/2$ or less.
\begin{figure}[t]
   \includegraphics[width=0.85\columnwidth]{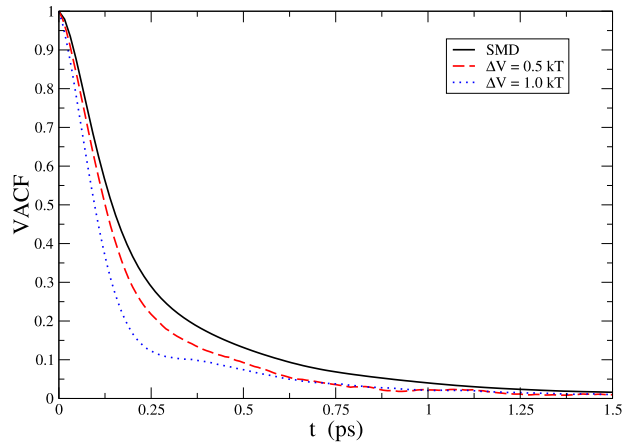}
   \caption{The normalized center of mass velocity autocorrelation
    function for different levels of energy terracing.}
\label{acfFig}
\end{figure}

Dynamical correlations, on the other hand, tend to be more sensitive
to the level of discretization in the stepped-potential. 
Considering the
normalized autocorrelation function of the velocity of the center of
mass (VACF), shown in Fig.~\ref{acfFig}, it is clear that the VACF
decays too quickly for large steps.
This behavior can be understood by noting that the average number of neighbors in the
first solvation shell is too large when $\Delta V = kT$, as can be
seen from the radial distribution functions in Fig.~\ref{rdfFig}.  
The increase in the average number of neighbors around a given particle 
leads to an enhanced number of interactions at short times that give
an exaggerated kick in random directions to the center of mass velocities and leads to
a rapid loss of correlation.
However as the
energy steps are reduced, there is a convergence of the VACF towards
the result of the SMD simulations since the local structure as well as the 
magnitude of the impulses are modified.  Note that when $\Delta V = kT/2$, the
life-time of velocity correlations 
is roughly correct. 

The same trends are observed in the orientational correlation
functions.  As is clear from Fig.~\ref{orientFig}, the normalized
autocorrelation function of the long-axis of the rods decays too slowly
when the energy discontinuities are too large.  Once again, the long
lifetime of orientational correlations for the system with large
discontinuities is indicative of pairs of molecules being too tightly
bound in an orientationally dependent, locally-preferred configuration.
However, the DMD results rapidly approach the SMD result as $\Delta V$
is reduced to $kT/2$ and exhibit the correct degree of anti-correlation 
for times around $0.15$ ps.

\begin{figure}[t]
   \includegraphics[width=0.85\columnwidth]{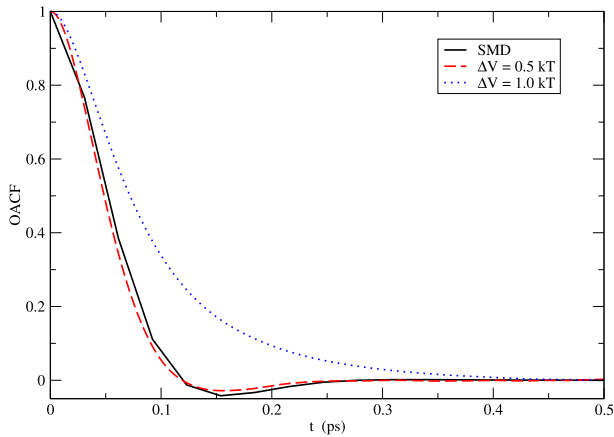}
   \caption{The normalized autocorrelation function of the long axis vector
   for different levels of energy terracing.}
\label{orientFig}
\end{figure}

For dense systems, the event-driven simulations are not expected to be
much more efficient than SMD simulations that take of advantage of stable and accurate
symplectic integration schemes.  The cost of the DMD simulation scales
linearly with the number of events processed per picosecond of time.
Interestingly, for $\Delta V$ in the range of $kT$ to $2$ $kT$, the
number of interactions per picosecond is approximately constant and
equal to $3.5 \times 10^4$ coll/ps.  At this rate of events, the DMD
simulations are roughly $1.7$ times more efficient than the SMD simulations.  However for
$\Delta V = kT/2$, where the dynamics compares well with the dynamics
of the continuous system, the rate of interactions per picosecond
increases to around $4.7\times10^4$ coll/ps and the relative efficiency of the
DMD simulations drops to $1.25$.
Of course these results depend strongly on a number
of conditions, including the physical conditions of simulation, and are
particularly sensitive to the density.  For example, 
for a gaseous system where $T^* = 10$ and $\rho^* = 0.01$, the DMD
simulation with $\Delta V = kT/2$ becomes more than $120$ times more efficient than the SMD simulations.
From an algorithmic point of
view, the low density DMD simulations resemble an adaptive time step
simulation approach in which large time steps are utilized to
integrate the equation of motion of a given particle when it is not
interacting with any others, and when another particle is encountered,
the time step is reduced to properly evolve the system through the
interaction region.  The energy-terracing scheme and event detection
algorithm naturally provide an adaptive approach in which an
interacting pair may experience many events in rapid succession as the
pair evolves through an interaction region.

\section{Conclusions}
\label{Conclusions}

In this article a general framework for performing event-driven
dynamics of rigid bodies has been presented that makes use of a
mapping of a continuous potential onto discrete values.  The evolution
of the discretized-energy system consists of periods of free
propagation of the system punctuated by impulses generated at discrete
times that correspond to moments when the underlying continuous
interaction potential for a pair of particles hits a critical value.
An adaptive grid method using cubic interpolation was presented to
facilitate the search for the earliest interaction time.  The effect
of the impulses on the subsequent dynamics of the system was analyzed
using conservation conditions, resulting in a simulation algorithm
that solves the evolution of the system within numerical precision.
The trajectories are time-reversible, and exactly obey all applicable
conservation laws by construction.

The method was demonstrated on a system of dense rigid rods
interacting by a discretized version of a simple modified
Lennard-Jones pair interaction potential in which the effective
distance depends explicitly on the relative orientation of the rods.
Although static correlations were reasonably well represented in the
discretized potential system provided the energy mapping was not too
coarse, dynamical correlations and related quantities such as the
orientational relaxation time depend sensitively on the level of
discretization of the continuous potential.  It was found that
potential energy steps on the order of $kT/2$ were required to
reproduce results from simulations of the continuous potential system.

It should be emphasized that the implementation tested here, in which
an evenly-spaced energy discretization was used, is the simplest
choice of mapping of the continuous system onto a set of discrete
energy levels.  It is quite likely that some other distribution of
energy levels would lead to substantial improvements in both the
quality of the simulation results and the overall efficiency of the
algorithm. However such modifications to the implementation must be
carefully considered, since typically one is only interested in rough
qualitative behavior of a system that is not overly sensitive to
details of an interaction potential.  In this case, the goal is to
construct a simple model that demonstrates the relevant physics.  
A few, well-chosen potential energy discontinuities can be expected to
meet this goal.

\section*{Acknowledgments}

The authors would like to acknowledge support by grants from the
Natural Sciences and Engineering Research Council of Canada (NSERC).


\begin{thebibliography}{99}
\bibitem{fna}
This symmetry is often broken in molecular simulations due to
periodic boundary conditions.
\bibitem{fnb} Another way to show that the scalars $S_i^l$ have to be
  equal, is to note that the direction of the impulse in six
  dimensional configuration space (including relative orientation and
  position) is in the direction of the normal to the boundary of the
  terrace being hit.  This normal is given by the gradient of the
  smooth underlying potential, and thus the six dimensional impulse
  has to have the same direction, with only a single scalar $S$ to be
  determined.
\end{thebibliography}
\end{document}